\documentclass[runningheads]{llncs}
\usepackage[T1]{fontenc}
\usepackage{comment}
\usepackage{orcidlink}
\usepackage[misc]{ifsym}
\usepackage{fancyhdr}
\usepackage[
    type={CC},
    modifier={by},
    version={4.0},
]{doclicense}
\usepackage{graphicx}

\usepackage{csquotes}
\usepackage[capitalise]{cleveref}
\usepackage{todonotes}
\definecolor{my-dark-red}{RGB}{183, 28, 28}
\definecolor{my-red}{RGB}{244,67,54}
\definecolor{my-pink}{RGB}{233,30,99}
\definecolor{my-purple}{RGB}{156,39,176}
\definecolor{my-deep-purple}{RGB}{103,58,183}
\definecolor{my-indigo}{RGB}{63,81,181}
\definecolor{my-blue}{RGB}{33,150,243}
\definecolor{my-light-blue}{RGB}{3,169,244}
\definecolor{my-cyan}{RGB}{0,188,212}
\definecolor{my-teal}{RGB}{0,150,136}
\definecolor{my-green}{RGB}{76,175,80}
\definecolor{my-light-green}{RGB}{139,195,74}
\definecolor{my-lime}{RGB}{205,220,57}
\definecolor{my-yellow}{RGB}{255,235,59}
\definecolor{my-amber}{RGB}{255,193,7}
\definecolor{my-orange}{RGB}{255,152,0}
\definecolor{my-deep-orange}{RGB}{255,87,34}
\definecolor{my-brown}{RGB}{121,85,72}
\definecolor{my-grey}{RGB}{158,158,158}
\definecolor{my-blue-grey}{RGB}{96,125,139}
\definecolor{my-lipics-grey}{rgb}{0.6,0.6,0.61}
\usepackage{listings}
\usepackage[scaled=0.85]{beramono}
\newcommand{\Cpp}{C++}

\usepackage{lmodern}
\usepackage[%
activate={true,nocompatibility},%
final,%
tracking=true,%
kerning=true,%
spacing=true,%
stretch=10,%
shrink=30]{microtype}

\usepackage{hyperref}
\title{Concepts for Designing Modern\\ \Cpp{} Interfaces for MPI}

\author{
C. Nicole Avans\inst{1}\orcidlink{0009-0005-0768-4243} \and
Alfredo A. Correa\inst{2}\orcidlink{0000-0002-9718-7099} \and
Sayan Ghosh\inst{3}\orcidlink{0000-0001-8758-7657} \and
Matthias Schimek\inst{4}\orcidlink{0009-0002-6402-9016} \and
Joseph Schuchart\inst{5}\orcidlink{0000-0003-2041-7877} \and
Anthony Skjellum\inst{1}\orcidlink{0000-0001-5252-6600} \and
Evan D\hbox{.} Suggs\inst{1}\orcidlink{0000-0002-8210-8992} \and
\hbox{Tim Niklas Uhl\inst{4}\textsuperscript{(\href{mailto:uhl@kit.edu}{\color{black}\Letter})}\orcidlink{0000-0001-9295-1388}}
}
\authorrunning{C\hbox{.} N\hbox{.} Avans et al.}
\institute{
Tennessee Technological University, Cookeville, Tennessee, USA\\ \email{\{cnavans42,askjellum,esuggs\}@tntech.edu}
\and
Lawrence Livermore National Laboratory, Livermore, California, USA\\ \email{correaa@llnl.gov}
\and
Pacific Northwest National Laboratory, Richland, Washington, USA\\ \email{sg0@pnnl.gov}
\and
Karlsruhe Institute of Technology, Karlsruhe, Germany\\ \email{\{schimek,uhl\}@kit.edu}
\and
Stony Brook University, Stony Brook, New York, USA\\ \email{joseph.schuchart@stonybrook.edu}
}
\fancypagestyle{firstpage}{
  \fancyhf{}
  \fancyfoot[L]{
    \footnotesize
    \copyright\; The Author(s) 2025 \\
    This version of the contribution has been accepted for publication after peer review, but is not the Version of Record and does not reflect post-acceptance improvements, or any corrections.
    The Version of Record is available online at: \url{https://dx.doi.org/10.1007/978-3-032-07194-1_10}
  }
}
\makeatletter
\def\sectionmark#1{}%
\def\subsectionmark#1{}%
\makeatother

\hypersetup{
  colorlinks=true,
  allcolors=blue,
  bookmarksdepth=2
}

\begin{document}
\pagestyle{fancy}
\maketitle              %
\thispagestyle{firstpage}
\setcounter{footnote}{0}

\begin{abstract}
    Since the C++ bindings were deleted in 2008, the Message Passing Interface (MPI) community has recently revived efforts in building high-level modern C++ interfaces.
    Such interfaces are either built to serve specific scientific application needs (with limited coverage to the underlying MPI functionality), or as an exercise in general-purpose programming model building, with the hope that bespoke interfaces can be broadly adopted to construct a variety of distributed-memory scientific applications.
    However, with the advent of modern C++-based heterogeneous programming models, GPUs and widespread Machine Learning (ML) usage in contemporary scientific computing, the role of prospective community-standardized high-level C++ interfaces to MPI is evolving.
    The success of such an interface clearly will depend on providing robust abstractions and features adhering to the generic programming principles that underpin the C++ programming language, without compromising on either performance or portability, the core principles upon which MPI was founded. However, there is a tension between idiomatic C++ handling of types and lifetimes and MPI's loose interpretation of object lifetimes\slash ownership and insistence on maintaining global states. 
    
    Instead of proposing ``yet another'' high-level  C++ interface to MPI, overlooking or providing partial solutions to work around the key issues concerning the dissonance between MPI semantics and idiomatic C++, this paper focuses on the three fundamental aspects of a high-level interface: type system, object lifetimes, and communication buffers, while also identifying inconsistencies in the MPI specification.
    Presumptive solutions can be unrefined, and we hope the broader MPI and C++ communities will engage with us in productive exchange of ideas and concerns. 

\end{abstract}

\section{Introduction }\label{sec:intro} %
Many modern C++ projects rely on MPI extensively, but MPI cannot natively handle most C++ data structures or constructs~\cite{mpi50}.
We plan to derive a basic C++ interface that takes into account new features and capabilities of C++ without compromising either performance or portability.
This paper seeks to derive key concepts and design considerations to promote the creation of a full modern MPI C++ interface, while highlighting issues complicating this endeavor.

C++ bindings were included in the MPI standard, but were deprecated by MPI 2.2 \cite{MPI2.2Spec}. 
These bindings have been entirely removed in version 3.0 (2012), since they only added minimal functionality over the C bindings while adding significant maintenance complexity to the MPI specification~\cite{Forum2009}.
C++ has changed significantly since the deprecation of C++ bindings in 2009, requiring a new look at its capabilities and the feasibility of a MPI C++ interface. %

 There are no guidelines now for developing efficient modern C++ interfaces over MPI (it is impossible to successfully enforce recipes such as C++ Core Guidelines~\cite{CppCoreGuidelines} without standardizing the behavior of underlying MPI objects, which is usually left up to concrete implementations).
 Here, we expand on some general considerations for designing C++ interfaces over MPI in good faith, without imposing ad hoc rules that might limit productivity, performance, and/or portability.

\noindent \textbf{Build on existing ideas.} We consider, re-evaluate, formalize and extend upon ideas presented in existing language bindings (\cref{sec:related-work}).

\noindent \textbf{Derive a C++ representation of MPI's object model.} We discuss how to bridge the gap between MPI objects and \Cpp{}, particularly how to handle their life cycle and formalize mutability in terms of constness (\cref{sec:object-model}).

\noindent \textbf{Enabling type safety in MPI.} The MPI C API currently does not utilize type information provided by the language. In \cref{sec:type-model}, we categorize \Cpp{} types and how these different classes can support type safe compatibility with MPI.

\noindent \textbf{Contiguous ranges as first class communication buffers.} \Cref{sec:data-buffers} describes rules defined through \Cpp{}-20 concepts, which specify which kind of \Cpp{} containers are directly supported by MPI for communication operations. This enables direct support for many STL containers and provides a powerful interface to derive more complex abstractions. We also describe how to handle data ownership, which is necessary to conform to the best practices of resource management in modern C++ (namely, RAII) in \cref{sec:ownership}. 

\noindent \textbf{Idiomatic error handling.} We describe how \Cpp{} can enhance MPI error handling via compile-time checking and exceptions in \cref{sec:error-handling}.

These are the major aspects and principles to which a future \Cpp{} MPI interface should adhere. 
Rather than proposing a complete standardized interface, which risks repeating the shortcomings of the removed bindings, our primary goal is to define \emph{semantic guidelines} and \emph{core conceptual interfaces} that align idiomatic \Cpp{} with MPI. 
These concepts could serve as the foundation for a MPI standard side document.
Standardizing concepts rather than concrete APIs offers the advantage of long-term maintainability and adaptability as both MPI and the \Cpp{} language evolve.
A \Cpp{} interface could introduce additional quality-of-life features to enhance productivity and provide safer abstraction, but these often introduce overheads that conflict with MPI's performance-portability while increasing the complexity of a specification.
Still, we outline major ideas and provide suggestions on how they could be aligned to our previously introduced design considerations. Particularly, we focus on serialization of more complex data types (\cref{sec:serialization}), first steps towards automatic life cycle handling of more complex MPI data types with direct mappings to \Cpp{} (\cref{sec:type-pool}) and how to increase MPI programmer's productivity by a sane set of defaults (\cref{sec:defaults}).

In \cref{sec:conclusions}, we conclude our discussion with suggestions on how the MPI standard could support language interface designers (not limited to \Cpp{}) by carefully extending the MPI specification. %

\section{Existing C++ Interfaces over MPI}
\label{sec:related-work}
Since the %
removal
of the \enquote{official} \Cpp{} MPI interface, a large number of third-party library interfaces have emerged, joined by efforts to bridge the gap between MPI implementations, \Cpp{}, and high-level performance portability frameworks (e.g., Kokkos). 
We will briefly outline notable works and their distinctive features in the following, as they serve as inspirations and proofs of concept for the technical discussion in the rest of this paper.

The \texttt{mpl} library~\cite{Bauke2015} is a C++-17 based, header-only library meant to provide an easy to use MPI interface for C++
developers after the deprecation and removal of the C++ API in MPI 3.0.
Recent work~\cite{ghosh2021towards} have extracted communication-specific interfaces from the main \texttt{mpl} codebase (reducing about 4K LoC in main \texttt{mpl}), and consider it as a prototypical modern C++ interface for studying MPI-specific language bindings.
To support bulk communication, \texttt{mpl} uses built-in \texttt{mpl::layout} class to manage derived data types.

Boost.MPI \cite{Gregor2007} offers a near one-to-one mapping of MPI-1 via free functions using communicators. 
Dynamic types are handled as skeletons containing address info, which must be updated if data is relocated (e.g., invalidated iterators).

B-MPI3~\cite{Correa2018} wraps MPI-3 with a C++ interface based on communicators and member functions, emphasizing const-correctness and iterator-based ranges.
It uses compile-time strategies to select communication methods: 
contiguous ranges use direct C-MPI calls; non-contiguous ones are copied; unsupported types are serialized via Boost.Serialization. 

RWTH-MPI~\cite{Demiralp2023} is a C++ interface for MPI that supports contiguous STL containers as send/receive buffers and offers overloaded MPI procedures with automatic parameter inference.
For custom types, it can auto-generate MPI data types using the PFR library \cite{Polukhin2016}.
Dynamic-size types are unsupported. While it covers MPI 4.0, its bindings largely mirror the C interface with limited added abstraction or safety.

KaMPIng~\cite{DBLP:conf/sc/UhlSHHKSS024} uses modern C++ features like move semantics to enhance safety, enabling return-by-value and memory-safe non-blocking communication.
It constructs MPI data types automatically for STL and custom types when possible and supports compile-time C++ to MPI type mapping.
Its named parameter interface with compile-time defaults facilitates both high-level prototyping and low-level control, keeping the full MPI interface accessible.

The Enhanced Message Passing Interface (EMPI)~\cite{beni2023empi} is based on modern C++, which is built on top of a customized version of Open MPI, eliding runtime checking overheads to directly map EMPI objects to low-level MPI objects within the Open MPI implementation. To enable RAII, EMPI proposes a \emph{program context} that wraps the MPI environment, and specializes MPI group from the context for abstracting communication and synchronization.

Kokkos Comm~\cite{KokkosComm10740805} utilizes modern C++ language features to provide more intuitive support of inter-node communication of Kokkos Views. Templates provide the needed flexibility to adapt to diverse data layouts and memory spaces.

MPI Advance~\cite{bienz2023mpiadvanceopensource} provides lightweight libraries that complement available system MPI installations to leverage tuned performance while also implementing support for the newest features from the MPI standard (e.g., partitioned communication) and additional capabilities and optimizations beyond the current scope of the standard.

\section{High-Level Design Considerations}\label{sec:components}
Existing work on designing C++ interfaces to MPI has brought up a huge amount of interesting design concepts and ideas, but most of the time they are hidden behind the details of the actual implementation. 
In this section, we will take a step back and clearly define the underlying semantic concepts of modern and idiomatic \Cpp{} MPI interface and how MPI's design can be made \enquote{compatible} with the \Cpp{} language. 
These concepts can then be used by implementers to design a concrete interface.

We start by defining how MPI objects map to \Cpp{} (\cref{sec:object-model}). Then we focus on how to model the actual data involved in communication, in terms of types (\cref{sec:type-model}) and the actual memory involved (\cref{sec:data-buffers}).
We then discuss how combining \Cpp{}'s ownership model and MPI to obtain additional memory safety (\cref{sec:ownership}) and how to handle errors in an idiomatic way (\cref{sec:error-handling}).

\subsection{Mapping the MPI Object Model to C++ } %
\label{sec:object-model}

MPI introduces a range of \emph{MPI objects}, such as communicators, data types, and requests, which are represented in the C API using \emph{opaque handles}. 
While this is the only viable approach in C, it hides key semantic properties of these objects, including ownership, lifetime, and identity.
In contrast, \Cpp{} offers native language features that allow these properties to be expressed explicitly and safely.

For more than a decade, \Cpp{} MPI library designers have advocated for representing MPI objects as first-class \Cpp{} objects that act as proxies for the implementation-defined objects behind the handles~\cite{Skjellum2001}. 

It is worth noting that most MPI procedures operating on these objects can be expressed either as member functions or as free functions. Both choices are equally valid, and this decision is largely independent of the conceptual model discussed here. In this work, we present communication operations as member functions, without advocating for one approach over the other.%

Modern \Cpp{} idioms---particularly RAII (Resource Acquisition Is Initialization)~\cite[16.5]{Stroustrup1994}\cite[E.6]{CppCoreGuidelines} and move semantics---naturally support resource management: MPI objects can be constructed via \texttt{MPI\_*\_create} in class constructors and released via \texttt{MPI\_*\_free} in destructors, with ownership safely transferable through move operations.
However, the original MPI design deviates from this model because of its reliance on global state and implicitly managed global objects---an approach discouraged by the \Cpp{} Core Guidelines~\cite[I.3,I.22]{CppCoreGuidelines}.
For example, global communicators such as \texttt{MPI\_COMM\_WORLD} are automatically initialized via \texttt{MPI\_Init} and cannot be explicitly created or freed.
Further, MPI prohibits reinitialization after \texttt{MPI\_Finalize}, limiting modular or library-based usage patterns.

\subsubsection{Sessions, Groups, and Communicators}
MPI 4.0 addressed  some of these limitations by introducing the \emph{Session Model}, an alternative to the traditional World Model for process management. Although originally designed to support better isolation in multi-threaded and multi-component environments, the session model aligns well with modern \Cpp{} object-oriented design. In this model, no global communicators are predefined; instead, communicators must be explicitly created from process sets via user code. This makes ownership explicit and fully under user control, enabling clean integration with RAII-based designs as shown in \cref{fig:object-model}. For example, communicators and groups are constructed and destroyed as part of their enclosing scope.

\begin{figure}
\begin{lstlisting}
mpi::session session{};
mpi::group group = session.group_from_pset("mpi://WORLD");
mpi::communicator comm{group}; // create communicator from group
comm.send(...);
\end{lstlisting}
\caption{Example of the proposed object model using the MPI Session Model}
\label{fig:object-model}
\end{figure}

\subsubsection{\texttt{const}-Correctness of MPI Objects}
The concept of constness is important for both the compiler and, more crucially, libraries, as it indicates variable mutability and restricts modifications in certain contexts. 
This applies straightforwardly  to the data being communicated: 
For instance, a receiving buffer cannot be \texttt{const}, which already enhances the program's safety, which we discuss in detail in \cref{sec:data-buffers}. 
In a less obvious way, constness can also be applied to library objects of a MPI C++ interface, such as communicators, data types, and requests and procedure calls on them, making it even more important to discuss here.

When creating a C++ wrapper for MPI or any C interface that does not account for constness, adding the keyword \texttt{const} requires a good grasp of the interface semantics and precise knowledge of the implementation (internal mutation).
Without explicit guarantees from the MPI standard specification, it is difficult to gather sufficient information to determine whether a particular operation can be marked as \texttt{const}.

We want to illustrate this for MPI communicators.
Most procedure calls leave a communicator in the same state as before, which would \emph{naively} allow them to be specified as \texttt{const}.
Conversely, one could argue that posting a message is not the same as not having posted it; in the former case, the message can be received on the other side, while in the latter, a receive operation may hang.
We think this is a strong argument to mark communication calls as non-const.
But this can still be challenged for procedures that send and receive in a single call (such as \texttt{send\_recv} or collectives).
\footnote{Note that this is a fundamental problem, not exclusive to const-qualified member functions; if the procedures are free functions the questions will still stand regarding the constness of the communicator argument parameter.}

To resolve these cases, we invoke a modern interpretation (post \Cpp11):
Since the communicator is likely undergoing internal mutation during any non-trivial operation, it is reasonable to conclude that most communication procedures should be marked as non-\texttt{const} at the communicator level.

\begin{figure}
\begin{lstlisting}[emph={}]
class mpi::communicator {
   communicator(communicator const& other) = delete;  // no copy-constructor
   ...
   auto duplicate() /*non-const*/ -> communicator;
   ...
   auto send(/*const data*/ ...) /*non-const*/;
   auto receive(/*mutable data*/ ...) /*non-const*/;
   auto broadcast(/*mutable data*/ ...) /*non-const*/;
   ...
   auto size() const;  // most likely can be marked const
                       // in a reasonable implementation
};
\end{lstlisting}
\caption{Interpretation of const-ness in a MPI \Cpp{} interface}
\label{fig:bmpi3-constness}
\end{figure}

If we adhere to the conclusion that communication procedures should not be marked \texttt{const}, an important and somewhat unexpected result follows: a communicator class should \emph{not} have a duplicate mechanism that is const on the original communicator.
Incidentally, there will be no canonical copy constructor (taking original communicator as a constant reference).
This is consistent, since even communicators duplicated from each other cannot receive or complete operations initiated in another communicator.
Thus, two communicators are never copies of each other; they are at most alternative virtual fabrics for subsequent communication.
\cref{fig:bmpi3-constness} summarizes our discussions about the communicator class which also align with the interpretation of the B.MPI3~\cite{Correa2018} interface.

For other parts of the library, the situation might be simpler. For example, most datatype manipulation is  likely to take advantage of \texttt{const}, since datatype manipulation usually generates new types instances from other types instances.

In summary, a key problem is that the MPI standard says little about the mutation of MPI \enquote{objects} (communicator, data types or request objects), which is further complicated by the existence of a global mutable state at the level of the environment.
In most cases, the internal mutation seems only to be implied by common knowledge, which makes the decision on the correct usage of \texttt{const}-ness in a \Cpp{} interface difficult.

Design and conventions on this issue, have important ripple effect in the design of C++ MPI programs, specifically C++ classes that contain communicators and other MPI objects, which is common in programs written at a high level of abstraction \cite{Andrade2021,Godoy2025}.
Classes that use their own internal communicator even for non-mutating operations would require mutation of the communicator, internally at least (i.e., \lstinline{mutable} attribute, and possibly synchronization).

\subsection{Modeling and Mapping Types}%
\label{sec:type-model}
Applications using MPI use a variety of data types that need to be communicated. The MPI standard distinguishes between \emph{basic datatypes} and arbitrarily complex \emph{derived datatypes}, which can be recursively constructed from other data types using type constructors (\texttt{MPI\_Type\_create\_*}).
C's lack of type introspection features forces users to always pass the type explicitly to a communication call, which is both tedious and error-prone, since type definitions need to be kept in sync with the actual data layout.
Fortunately, for many C++ types, there is a one-to-one mapping to MPI data types. C++ defines a set of \emph{fundamental types}: 
\texttt{void}, \texttt{std::nullptr\_t}, \emph{integral types} (including integers, character types and \texttt{bool}) and \emph{floating-point types}. 
For integral and floating-point types, there exist matching predefined \emph{basic data types} in MPI.

Using template-metaprogramming, a C++ MPI interface can therefore deduce an MPI type directly from a data buffer (as defined in \cref{sec:data-buffers}), in case its underlying \texttt{value\_type} is fundamental.
This approach is implemented by all major MPI C++ bindings~\cite{Bauke2015,Correa2018,Gregor2007,DBLP:conf/sc/UhlSHHKSS024}.
More complex types require explicit creation and a subsequent commit step, and have to be freed before MPI is finalized. To enable proper cleanup we again use RAII and represent data types as \texttt{mpi::datatype} objects, which support move construction and assignment and free the data type when the destructor is called.

To prevent users from using uncommitted data types in communication, we propose to encode the commit information as part of the type, similar to the approach of \emph{rsmpi}~\cite{Steinbusch2015}. The function \texttt{mpi::commit} takes an rvalue \texttt{mpi::datatype} object, and converts it to a \texttt{mpi::committed\_datatype} as shown in \cref{fig:commit}. Accordingly, communication calls only accept data type objects of this type.

\begin{figure}
\begin{lstlisting}
struct MyType {
  int a;
  std::array<int, 3> b;   
  double c;
  char d;
};
mpi::datatype type = mpi::datatype::for<MyType>();
mpi::committed_datatype struct_type = mpi::commit(std::move(type));
\end{lstlisting}

\caption{Example for constructing a type for a C++ type and committing it}
\label{fig:commit}
\end{figure}
 
In addition to fundamental C++ types with direct mappings to predefined MPI basic types, there exists a broader class of C++ types that can still be safely mapped to MPI data types.
These are precisely the types classified \emph{trivially copyable} by the \Cpp{} standard.
A type is trivially copyable if its binary representation can be safely copied byte-by-byte (e.g., via \texttt{memcpy}), without violating language rules and invoking undefined behavior. More concretely, an object of such type can be copied to an array of \texttt{char}, \texttt{unsigned char}, or \texttt{std::byte}, and then back into another object of the same type, which will hold the same value as the original.
This property is essential for MPI communication: when transmitting data across processes, the memory content of a variable must be sent as a sequence of raw bytes over the network and reconstructed correctly on the receiving end. 
If the type is not trivially copyable, this round-trip may not preserve the original value or may result in undefined behavior. Thus, restricting to trivially copyable types ensures that the MPI type correctly matches the actual in-memory layout and semantics of the transmitted data.

These types can be safely supported by automatically constructing the corresponding MPI datatype using a compile-time reflection mechanism to call the correct MPI type constructors. This ensures that the \Cpp{} type is compatible with MPI's type system and enables safe type handling, even on heterogeneous systems.
Although C++ currently lacks native language support for such reflection, concrete proposals are in place to introduce compile-time type reflection in C++26 \cite{Revzin2024P2996R5}. 
In the meantime, third-party libraries such as Boost.PFR \cite{Polukhin2016} can be used to implement this feature, although they have some limitations, for example for types using inheritance or private members.

Beyond fundamental and trivially copyable types, there remains a broad class of other (potentially user-defined) \Cpp{} types that do not have a direct or automatically derivable mapping to an MPI data type. 
For these types, the correspondence between memory layout and type semantics cannot be safely inferred by the prospective \Cpp{} MPI interface. 
Consequently, it becomes the user's responsibility to explicitly define how the type is laid out in memory and how that layout corresponds to an appropriate MPI datatype.
While the handling of fundamental and trivially copyable types can be automated to reduce programming errors and ease MPI development, a prerequisite for such automation is a clear conceptual model of the data involved in communication.
The core abstraction in this model is the \emph{data buffer}, which we introduce in the following \cref{sec:data-buffers}.

\subsection{Modeling Memory Involved in MPI Communication } %
\label{sec:data-buffers}
We now explore how a C++ MPI interface can offer idiomatic abstractions for the data sent or received in MPI communication operations.  
In MPI, this data is described in terms of a pointer to a memory region, an (MPI) datatype, the number of elements to be sent, and (for some collective operations) their displacements.
While this number of parameters makes MPI  flexible, it also leads to verbose function calls.
Further, the flexibility of MPI is only required in a few use cases and results in unnecessarily complicated code in other cases~\cite{DBLP:conf/sc/UhlSHHKSS024}.

Particularly when using standard library containers such as \texttt{std::vector<T>} with the MPI C API, users have to access underlying raw pointers and sizes of the containers and pass them to MPI individually, while the actual vector object already comprises the whole send data context.
This is both non-ergonomic to use and conflicts with the \Cpp{} Core Guidelines~\cite[I.13]{CppCoreGuidelines}. Most existing MPI C++ libraries therefore introduce support for a subset of standard library containers~\cite{Bauke2015,Correa2018,Demiralp2023,Gregor2007,DBLP:conf/sc/UhlSHHKSS024} as their core abstraction feature.
However, this has certain shortcomings.
First, this is often defined ad-hoc only including a fixed set of containers~\cite{Demiralp2023,Gregor2007}.
Second, design decisions within the approaches of supporting (standard library) containers introduce hidden, additional overheads through memory allocations or additional MPI calls~\cite{Bauke2015,Gregor2007,DBLP:conf/sc/UhlSHHKSS024}.
For example, in an \texttt{MPI\_Recv} call, Boost.MPI resizes the container into which it receives the data to match the size of the incoming message.
If the user does not want this, they must instead pass raw pointers again.
The KaMPIng library internally invokes \texttt{MPI\_Probe} when the user issues a receive operation without a \texttt{count} argument. 
Such unexpected overheads are not desired from the perspective of a standardized C++ interface.

Therefore, in the following, instead of reiterating the specific approaches of existing MPI wrappers and defining how a particular container is \enquote{fitted} to support MPI, we will define a small set of underlying rules and categories for memory involved in communication in terms of C++ \emph{concepts}.
Concepts are a (modern) C++ language feature for specifying constraints on (custom) types.
C++ provides easy-to-use mechanisms to check that types satisfy a given concept at compile time.
This gives us well-defined semantics on which kind of containers can be used directly with MPI, deducing size and type information safely from \Cpp{}, without imposing any additional overhead. Additionally, C++ objects not satisfying required concepts result in easy-to-understand error messages, as opposed to previous often very verbose and complicated error messages caused by template metaprogramming errors~\cite{cpp_constraints}.

For each send or receive buffer involved in a communication call, MPI's C API expects a pointer to a contiguous memory location, a count argument and an MPI data type.
\Cpp{} already provides type information, and automatically matching types is crucial for type safety. A high-level standardizable C++ interface, can use type introspection to provide such type safety, which we detail in \cref{sec:type-model}. For now, let us assume that we can always deduce an MPI data type from a provided container.
We call this basic building block of MPI communication (memory location, count and type) a \emph{data buffer}. %
In the following, we describe a minimal interface of data buffer objects in terms of \Cpp{} concepts. 

A data buffer object has to satisfy the concept \texttt{std::range}; that is, it must expose an iterator to the beginning (and to the end) of its underlying storage.
Additionally, data buffers must satisfy the following (built-in) concepts:
\begin{itemize}
	\item \texttt{std::ranges::contiguous\_range}, that is, its underlying memory has to be contiguous in memory. This captures MPI's requirement that data which is communicated must reside in \emph{sequential storage}~\footnote{An exception to this requirement is the usage of \texttt{MPI\_BOTTOM} which relies on absolute memory addresses to describe data and should be treated separately.}.
	\item \texttt{std::ranges::sized\_range}, that is, a data buffer exposes a \texttt{size()} function returning the number of its elements.
    \item \texttt{Typed}, that is, either the data buffer exposes a constant reference to a committed MPI data type or the data buffer's \texttt{value\_type} is a fundamental type (for which a direct mapping to predefined MPI data types exists and can directly be used if not overwritten by the user).
\end{itemize}

\begin{figure}[t]
\centering
	\begin{lstlisting}
namespace mpi {
  // Concept ensuring that t exposes MPI data type information
  template <typename T>
  concept Typed = requires(T t) {
      { mpi::datatype(t) } -> std::same_as<committed_datatype const&>;
  };
  
  template <typename T>
  concept DataBuffer =
      std::ranges::contiguous_range<T> &&
      std::ranges::sized_range<T> &&
      (std::is_fundamental_v<typename T::value_type> || Typed<T>);
      
  // Concept for data buffers storing send data
  template <typename T>
  concept SendDataBuffer =
      DataBuffer<T> &&
      std::ranges::input_range<T>;
  
  // Concept for data buffers encapsulating data to be received
  template <typename T>
  concept RecvDataBuffer =
      DataBuffer<T> &&
      std::ranges::output_range<T, typename T::value_type>;
}
\end{lstlisting}
	\caption{Concept definition for a data buffer}
	\label{fig:data-buffer-concept}
\end{figure}

An example of how these constraints can be expressed in plain \Cpp{} is given in~\cref{fig:data-buffer-concept}, which defines the concept \texttt{mpi::DataBuffer} in terms of the three concepts described above.
A  convenient property of this data buffer definition is that many C++ standard library containers like \texttt{std::vector} or \texttt{std::array} storing fundamental \Cpp{} types such as \texttt{char, int, double, ...} already satisfy the \texttt{mpi::DataBuffer} concept. This also holds for lightweight, non-owning views such as \texttt{std::span}.
Hence, they can be directly used in an MPI C++ interface adhering to these principles.

For more complex cases, for example, to communicate a container of a custom \Cpp{} datatype which is \texttt{trivially\_copyable}, we first have to construct and commit such a datatype (see \cref{sec:type-model}) and then pass it to the data buffer.
For communicating data using complex derived datatypes without a one-to-one mapping to the underlying \Cpp{} type, we not only have to construct and commit an MPI datatype but also pass the number of corresponding elements to the data buffer.
For these more complex cases, there is no direct C++ standard library support but an MPI C++ interface could easily provide a lightweight \emph{adapter} with multiple constructors not only accepting a contiguous \texttt{std::range}, but also a count and a \texttt{mpi::committed\_datatype} parameter.

This also makes this interface  extensible to support more complex containers from third-party libraries like \texttt{Kokkos::View} or \texttt{thrust::device\_vector}, backed by memory located on accelerators such as GPUs. Therefore the data buffer concept naturally extends to the notion of \emph{memory allocation kinds}~\cite{memory-allocation-kinds}, providing direct support for accelerator-aware MPI on C++ containers backed memory located on GPUs or other accelerators.

\Cref{fig:api-example} illustrates exemplary communication calls showcasing the different cases discussed above.
Note that we do not suggest a concrete API or function signatures but rather want to show the underlying concept of using data buffers as an abstract description for communication data.

Orthogonal to the \texttt{mpi::DataBuffer} concept presented so far, we additionally have to model that data buffers describing send data have to be \emph{readable} and data buffers for receive data have to be \emph{writable}.
This can be expressed using the (built-in) concepts, \texttt{std::ranges::in/output\_range} (see~\cref{fig:data-buffer-concept}).

To conclude, in many (simple) cases, this approach leads to clean and less convoluted code with direct C++ standard library integration.
For more complex cases, we end up with the same number of parameters as in the C API, as we need to leverage its full flexibility. 
However, in all scenarios, we gain clear semantics directly documented \emph{in} source code by using C++ concepts.

\begin{figure}
\centering
\begin{lstlisting}[escapechar=!]
std::vector<int> v = ...    
comm.send(v, /*additional parameters*/);

std::vector<MyType> v = ... // static custom type defined in !\textcolor{my-dark-red}{\cref{fig:commit}}!
mpi::datatype type = mpi::datatype::for<MyType>();
mpi::committed_datatype type_c = mpi::commit(std::move(type));
comm.send(mpi::buffer_adapter(v, type_c), /*additional parameters */);

std::vector<char> v = ...
int count = ...
mpi::datatype complex_type = ... // complex derived type manually defined by user
mpi::committed_datatype complex_type_c = mpi::commit(std::move(complex_type));
comm.send(mpi::buffer_adapter(v, complex_type_c, count), /*additional parameters */);

int local_count = ...;
comm.reduce(mpi::single_adapter(local_count), ...); // support for non-range single arguments
\end{lstlisting}
	\caption{Exemplary communication calls using the data buffer concept when MPI procedures are implemented as member functions.}
	\label{fig:api-example}
\end{figure}

\paragraph{Extending the Data Buffer Concept.}
In collective communication calls with varying counts (such as \texttt{MPI\_Alltoallv}), MPI requires not just a single \emph{count} parameter but a separate count for each MPI process (aka rank).
Additionally, displacements can be specified for each rank. %
To reflect these semantic changes, we have to extend the \texttt{mpi::DataBuffer} concept as previously defined (\cref{fig:data-buffer-concept}) and define \emph{irregular} \texttt{mpi::DataBuffers}.
These buffers are still based on the concept \texttt{std::ranges::contiguous} and have to expose MPI datatype information.
Additionally, they also have to expose \texttt{size\_v()} and \texttt{displacements()} functions, which return the respective counts and displacement data.
For brevity, we refrain from providing an example definition of this additional concept.
Again, simple lightweight adapters can be provided for standard library and similar custom containers. 
Similarly, scalar arguments (such as a single \texttt{int} or \texttt{bool} in a reduction) can be easily supported via lightweight adapters, which implicitly return \texttt{size} $1$ and implement the data buffer concept (\cref{fig:api-example}).

\subsection{Ownership and Non-Blocking Communication}
\label{sec:ownership}

Using the data buffer concept instead of raw pointers already offers benefits such as an ergonomic interaction with standard library containers, improved type safety and the potential to prevent out-of-bounds accesses.
Another important advantage of this object-oriented data handling is that it allows us to model a proper concept of ownership.
In C++, if an object \emph{owns} a resource---such as heap-allocated memory---the object is responsible for releasing this resource when it goes out of scope.
This is usually achieved by placing the clean-up code in the object's destructor and the key idea behind the RAII idiom.
A data buffer owning its underlying memory is therefore responsible for freeing this memory once it goes out of scope, thus preventing memory leaks. 
This plays  well with C++'s move semantics.
During move-construction or move-assignment, ownership of the underlying resource is transferred from the source to the destination object.
Using this mechanism it is for example possible to move a (receive) data buffer to an MPI call as shown in~\cref{fig:move-semantics}.
Then the MPI call receives data into the provided buffer, which it now owns.
Once the communication is completed, it returns the updated receive buffer by value resulting in a clean and idiomatic pass through of the underlying memory resource.
This data flow allows us to avoid the use of \emph{out} parameters that are discouraged in \Cpp{}~\cite[F.20]{CppCoreGuidelines}.

In terms of library interface design, this \enquote{pass through} of data can be achieved by considering the \emph{value category} of data buffers passed to communication calls. We propose that a \Cpp{} interface should explicitly distinguish between arguments passed as r- or l-values: If a data buffer is an \emph{r-value}, the data buffer is moved to the MPI operation, and returned to the caller as a return value of the function call. If a data buffer is an \emph{l-value}, it is passed by reference to the operation, ownership is not transferred, and it is not returned. For an example see \cref{fig:move-semantics}.

Further, (moving) ownership is particularly useful for enhancing the safety of non-blocking communication.
In MPI, a call to a non-blocking communication  operation \emph{initiates} the operation but does not \emph{complete} it.
Instead, it returns a request handle, which the user can test for completion via \texttt{MPI\_Test} or wait (blocking) on via \texttt{MPI\_Wait}.
This return of control between initiation and completion of the underlying call introduces a potential source of programming errors since MPI semantics require that any buffer involved in a non-blocking operation remain unmodified by the user until the operation has completed.

A robust solution to this problem is to transfer ownership of data buffers involved in non-blocking communication calls to the C++ interface, which forward it to request object as show in \cref{fig:move-semantics}. The request object is conceptually similar to \texttt{std::future}
\footnote{Using \texttt{std::future} to provide a safe interface for non-blocking communicating is not possible, as they are tied to asynchronous progress happening in the background, which the MPI standard does not guarantee.}:
Users can only access the data upon completing the request via \texttt{wait} or \texttt{test}, which move the data back to the caller, either directly by value in the former case, or encapsulated in \texttt{std::optional} in the latter. 
This way, invalid accesses to data involved in non-blocking communication can be prevented entirely through library semantics. This idea was first introduced and implemented in the MPI wrapper KaMPIng~\cite{DBLP:conf/sc/UhlSHHKSS024}.

\begin{figure}
\centering
	\begin{lstlisting}
std::vector<int> recv_buf = {...};
recv_buf = comm.recv(std::move(recv_buf), ...); // ownership transferred to call 
                                                // and back to caller
comm.recv(recv_buf, ...); // buffer only captured by reference, and nothing returned

// Non-blocking communication with ownership transfer
mpi::request<std::vector<int>> req = comm.irecv(std::move(recv_buf), ...);
recv_buf = req.wait(...);
std::optional<std::vector<int>> result = req.test(...);
\end{lstlisting}
	\caption{Transferring ownership through \Cpp's move semantics}
	\label{fig:move-semantics}
\end{figure}

\subsection{Error Handling } %
\label{sec:error-handling}
MPI notifies users of errors by returning error codes from almost every function defined by the MPI standard.
Although this is the most common way to handle errors in C, this does not fit well with modern \Cpp{}.
Most current MPI C++ interfaces either ignore errors completely or encapsulate the returned error code in an exception and throw it. This has a shortcoming: MPI makes no distinction between failures, which may be recoverable, such as insufficient buffer space or node failures (when using a ULFM-enabled MPI implementation~\cite{ulfm}) and usage errors, such as providing invalid parameters, which cannot be resolved. The strategy of converting all returned errors to exceptions is opposed to the \Cpp{} Core Guidelines~\cite{CppCoreGuidelines} which give the following suggestions on error handling:
    (E.2) \enquote{Throw an exception to signal that a function can’t perform its assigned task,} and
    (P.5) \enquote{Prefer compile-time checking to run-time checking.}
Guided by this, we propose the following strategy (implemented by KaMPIng~\cite{DBLP:conf/sc/UhlSHHKSS024} and B.MPI3~\cite{Correa2018}):

First, usage errors, such as invalid parameter combinations or invalid types, are handled at compile time via \texttt{static\_assert}.
Since C++ template meta-programming is notorious for complex, hard-to-read compiler errors, we try to ensure that compile-time assertions fail early and provide helpful human-readable error messages. 
Looking ahead, \Cpp{}-26's enhancements to \texttt{static\_assert} will enable us to supply constant-expression diagnostic messages, potentially leveraging \texttt{std::format}, so that compile-time checks can convey rich contextual information.

For invariants that can only be verified during execution, we use a layered assertion system: We verify invariants ranging from lightweight checks to assertions involving additional communication. These can be disabled level-by-level at compile-time,  encouraging developers to use exhaustive checking while writing and testing code, yet permitting a lean, high-performance configuration for production builds. By integrating error-handling controls directly into the C++ interface---rather than relying on rebuilding the MPI library itself in a \enquote{debug} mode---we provide an ergonomic, flexible framework. If an assertion is enabled and fails, we call \texttt{MPI\_Abort} because execution can not safely continue.

Errors that may be recoverable are signaled to the user.
This can either by done by throwing an exception, or by returning \texttt{std::expected} from MPI procedures.
\Cpp{}-23's \texttt{std::expected<T, E>} offers a lightweight, value-based alternative to exceptions, by encapsulating either a valid \texttt{T} or an error \texttt{E} in one object. 
Using \texttt{std::expected} makes error handling explicit, avoids hidden blocking operations inside distant \texttt{catch} handlers, and allows for easy error chaining and propagation.
Together with the return-by-value-based design proposed in \cref{sec:ownership}, this extends the concept of error codes in MPI's C interface, by only providing access to the return value of a procedure in case of success.

\section{Considerations Beyond a Preliminary C++ Interface}
In the following, we discuss additional features which improve the usability of preliminary \Cpp{} MPI language bindings. These features are perhaps beyond the scope of initial standardization, since they introduce additional overhead on top of the current MPI specification, such as allocation or communication under the hood. 
Nevertheless, they make working with MPI from \Cpp{} easier and often safer. In the following, we discuss how to easily communicate data not directly compatible with MPI (\cref{sec:serialization}), how to improve the handling of type lifetimes (\cref{sec:type-pool}) and to improve programmer productivity and reduce boilerplate code by offering sane defaults to programmers (\cref{sec:defaults}).
\subsection{Serialization } %
\label{sec:serialization}

The data buffer concept defined in \cref{sec:data-buffers} provides a type-safe and easy-to-use way to use standard C++ containers and user-defined types with a regular memory layout and a direct mapping to an MPI data type in MPI communication. 
Commonly used complex C++ types, such as \texttt{std::list}, \texttt{std::unordered\_map}, or polymorphic class hierarchies, do not provide contiguous storage or a well-defined memory representation.
These are \emph{dynamic types}: their size or layout may vary at runtime, they may require logic to (re)construct from memory (on both ends of the communication) and may involve intermediate allocations associated with data copies.
While such types cannot be communicated directly using the MPI type system, a user-defined \emph{data buffer} encapsulating packed data data can be used. %
However, constructing this is cumbersome and error-prone.

Instead, we propose a general serialization mechanism: The solution is to serialize complex types into a contiguous byte buffer (e.g., a \texttt{std::vector<char>}), which are directly transmittable via MPI.
The deserialization step on the receiving side reconstructs the original object from this representation.

Importantly, serialization libraries such as Boost.Serialization~\cite{Ramey2009} or Cereal~\cite{Grant2017} let users define what data members of a type should be serialized;  that is, what constitutes the object's logical state.
The concrete binary layout or packing is handled automatically by the library and is treated as an implementation detail
\footnote{Implementations may even avoid byte packing and communicate in chunks.}.
This cleanly separates the intent of serialization from its mechanics, which is especially useful for complex types.
Constructing such a serialized buffer may itself involve MPI communication, for example to exchange message sizes ahead of receiving the actual payload.
While this is beyond the scope of the MPI standard, it can be built naturally on top of the previously defined data buffer concept.
A user-defined type wrapping the serialization logic and exposing the required interface (contiguous, sized, and typed buffer) integrates seamlessly with the rest of the system.

In summary, even though dynamic types require custom logic and cannot be described via static type traits, the buffer abstraction is general enough to accommodate them.
Serialization turns arbitrary C++ objects into something communicable without burdening the MPI interface with additional complexity.

\subsection{Safer Usage of Complex Types}
\label{sec:type-pool}
The type classification and data buffer concept introduced in \cref{sec:type-model,sec:data-buffers} provide all necessary building blocks to work with complex types. Even if a type is not \emph{directly} supported by MPI, for the large class of trivially copyable types, there still exists a type-safe, direct type mapping. This allows the user to semi-automatically construct a type, and pass it alongside a container to the communication call as a data buffer, as shown in \cref{fig:api-example}.

But this has shortcomings: Committing an MPI data type is not free, since implementations may optimize the internal representation of a type when it is committed. So in a \Cpp{} interface, it is the user's responsibility to commit the type and free it when it is no longer needed (by destructing the type object using the RAII pattern). This introduces additional problems, since now the user has to keep the type around all the time, and the type correspondence between the container and the data type object is lost. To improve usability here, we suggest the idea of a \emph{type pool} that allows to store committed data types and to perform a lookup based on type information.

Users then just construct a data buffer from a container of a trivially copyable type and a reference to the type pool and pass it to the communication call.
The communication call can then perform a runtime lookup based on \texttt{std::type\_info::hash\_code} of the containers value type to obtain the MPI data type used in communication. 

There is one caveat to this approach: The lifetime of types in MPI is currently not clearly defined. In the session model, the lifetime of types is allowed to span across multiple sessions which suggests that the MPI type system exists orthogonally to MPI initialization. But it also mandates that types may only be created once a session or the world model is initialized, yet users are encouraged to commit types separately for each session.

This inconsistency appears to be unintentional but has not been resolved since sessions were introduced\footnote{\url{https://github.com/mpi-forum/mpi-issues/issues/733}}. Fixing this will enable language bindings to MPI, not only limited to \Cpp{}, to handle types more easily. For example, when type lifetime is tied to sessions, a \Cpp{} implementation can associate a session-local type pool with each session object, which can be queried for types without providing a pool explicitly. Types can then be freed automatically when the session is finalized.

\subsection{Improving MPI Productivity with Sensible Defaults }%
\label{sec:defaults}
Many libraries improve usability features beyond a high-level standardized interface: automatically resizing receive containers, computing missing counts or displacements in collectives with varying counts, or even performing additional communication to determine missing arguments. 
For example, Boost.MPI~\cite{Gregor2007} and RWTH-MPI~\cite{Demiralp2023} offer various overloads for communication calls with default arguments, preventing users from writing a lot of boilerplate code.

KaMPIng~\cite{DBLP:conf/sc/UhlSHHKSS024} goes even further: It chooses an alternative approach inspired by \emph{named parameters}, where parameters passed to a function can be named at the caller site and passed in arbitrary order (as seen in languages like Python).%
This allows to check for the presence of each parameter and to compute default values only if the respective parameter is omitted, without resorting to many overloads exploring the complete combinatorial explosion of parameters.
To avoid run-time overhead, they rely on template meta-programming to only generate the code paths required for computing missing parameters at compile time.

While this is well beyond the scope of a low-level interface due to additionally introduced allocations or communications, concrete implementations building on this interface could introduce such improvements to make writing MPI code easier, less error-prone and more productive, by handling  resources on behalf of the user.

\section{Conclusions }\label{sec:conclusions} %
Standardizing a modern \Cpp{} interface is an ambitious---and perhaps impractical---undertaking. 
Instead, we advocated here for the creation of language support guideline  documents, similar in spirit to what this work aims to contribute.

While many desire a truly modern C++ interface for MPI, this paper has laid out the 
conceptual and technical challenges that need to be addressed before such an interface can be standardized.
Our goal was to clarify these challenges and provide a foundation of design principles that could guide future development and community discussion.
To this end, we covered high-level considerations: a \Cpp representation of the MPI object model (\cref{sec:object-model}), data representation (\cref{sec:data-buffers}) and ownership (\cref{sec:ownership}), modeling and mapping types (\cref{sec:type-model}), and idiomatic error handling (\cref{sec:error-handling}). We also discussed ideas beyond these core concepts, including serialization (\cref{sec:serialization}) and usability improvements (\cref{sec:type-pool,sec:defaults}).

Based on these considerations, we identified specific aspects where the MPI standard 
requires clarification---most notably the lifecycle of user-defined data types, which 
remains under-specified in the context of sessions and makes automated resource 
management more difficult.
We also support ongoing efforts to improve attribute support across the standard, 
particularly for sessions and requests, as this is essential for associating 
higher-level interface constructs with the lifetime of MPI objects 
(see~\href{https://github.com/mpi-forum/mpi-issues/issues/664}{Issue \#664}). 
Finally, while callback support already exists in the standard, not all callbacks 
currently support user-defined state. Enabling this consistently would improve the 
design and usability of language bindings 
(see~\href{https://github.com/mpi-forum/mpi-issues/issues/839}{Issue \#839}).

We hope that our discussion can serve as a blueprint for other MPI language communities.
A prime example for language support is Vapaa~\cite{HammondVAPAA}, a standalone implementation of MPI's Fortran interface built on the C API, which strongly influenced the addition of an MPI Application Binary Interface in the MPI 5.0 standard~\cite{Hammond2023}.
Also, many of the mentioned concepts can be applied to or were inspired from the Rust MPI bindings \texttt{rsmpi}~\cite{Steinbusch2015}, since modern \Cpp{} and Rust share many conceptual similarities.
For other languages, it might be helpful to derive their own, independent set of guidelines.
This will enable MPI continue to evolve together with popular languages that define the future of high-performance computing.

\begin{credits}
\subsubsection*{\ackname}
PSAAP Funding in part is acknowledged from these NSF Grants OAC-2514054, CNS-2450093, CCF-2405142, and CCF-2412182 and the U.S. Department of Energy's National Nuclear Security Administration (NNSA) under the Predictive Science Academic Alliance Program (PSAAP-III), Award DE-NA0003966.
AAC work performed under the auspices of the US Department of Energy by Lawrence Livermore National  Laboratory under contract DE-AC52-07NA27344,
and supported by the Center for Non-Perturbative Studies of Functional Materials Under Non-Equilibrium Conditions (NPNEQ) funded by the Computational Materials Sciences Program of the US Department of Energy, Office of Science, Basic Energy Sciences, Materials Sciences and Engineering Division. This work was also performed under the auspices of the US Department of Energy's Pacific Northwest National Laboratory, operated by Battelle Memorial Institute under contract DE-AC05-76RL01830. 
Any opinions, findings, and conclusions or recommendations expressed in this material are those of the authors and do not necessarily reflect the views of the %
National Science Foundation, or the U.S.\hbox{} Department of Energy's National Nuclear Security Administration.

\noindent
\begin{minipage}[t]{0.77\linewidth}
 This project has received funding from the European Research Council (ERC)
 under the European Union’s Horizon 2020 research and innovation program
 (grant agreement No.\ 882500).
\end{minipage}%
\hspace{1em}
\noindent
\begin{minipage}[t]{.2\linewidth}
\raisebox{-2.25em}{%
    \includegraphics[width=\linewidth]{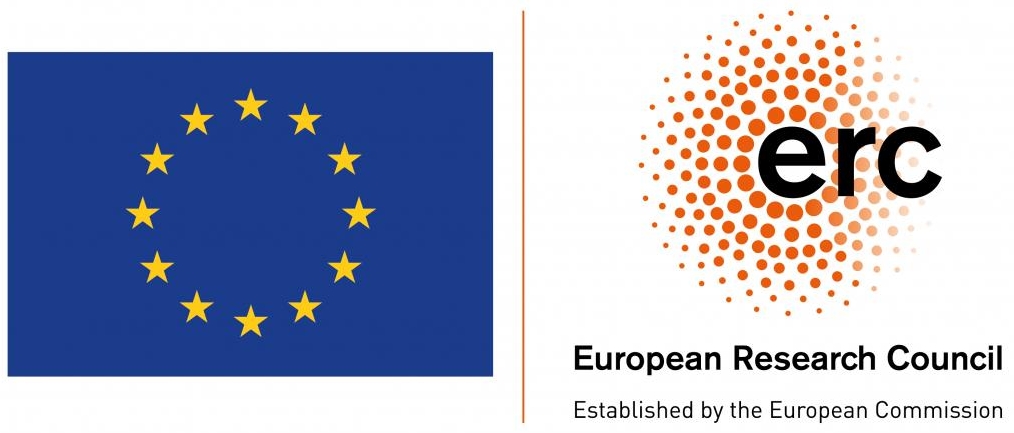}%
    }
\end{minipage}
\end{credits}
\newpage

\newpage
\bibliographystyle{splncs04}
\bibliography{paper}

\makeatletter
\setlength{\doclicense@hsize}{\linewidth-\doclicense@imagewidth-\doclicense@imagedistance}
\begin{center}
  \noindent\begin{minipage}{\doclicense@hsize}
    \subsubsection*{Open Access.}
    \noindent
    This chapter is licensed under the terms of the \doclicenseLongName License (\url{\doclicenseURL}),
    which permits use, sharing, adaptation, distribution and reproduction in any medium
    or format, as long as you give appropriate credit to the original author(s) and the
    source, provide a link to the Creative Commons license and indicate if changes were
    made.
  \end{minipage}
  \hfill
  \begin{minipage}{\doclicense@imagewidth}
    \doclicenseImage%
  \end{minipage}
\end{center}
\vspace{-.75em}

\indent The images or other third party material in this chapter are included in the
chapter’s Creative Commons license, unless indicated otherwise in a credit line to the
material. If material is not included in the chapter’s Creative Commons license and
your intended use is not permitted by statutory regulation or exceeds the permitted
use, you will need to obtain permission directly from the copyright holder.

\end{document}